\documentclass{svproc}
\usepackage{url}
\usepackage{xcolor}
\usepackage{graphicx}
\usepackage{multirow}
\usepackage{cite}
\usepackage{hyperref}
\usepackage{amsmath}
\usepackage{capt-of}
\usepackage[inline]{enumitem}
\usepackage{soul}
\usepackage{marvosym}

\newif\ifcomment

\begin{document}
\mainmatter              % start of a contribution
\title{Correlation analysis of node and edge centrality measures in artificial complex networks}
\titlerunning{Correlation analysis of centrality measures}  % abbreviated title (for running head)
\author{Annamaria Ficara\inst{1,2}$^{\textrm{\Letter}}$(0000-0001-9517-4131) \and Giacomo Fiumara\inst{2}(0000-0003-1528-7203) \and Pasquale De Meo\inst{2}(0000-0001-7421-216X) \and Antonio Liotta\inst{3}(0000-0002-2773-4421)
}
\authorrunning{Annamaria Ficara et al.} % abbreviated author list (for running head)
\tocauthor{Annamaria Ficara, Giacomo Fiumara, Pasquale De Meo and Antonio Liotta}
\institute{University of Palermo, Palermo, Italy,\\
\email{aficara@unime.it},
\and
University of Messina, Messina, Italy,\\
\email{\{gfiumara, pdemeo\}@unime.it},
\and
University of Bozen-Bolzano, Bolzano, Italy,\\
\email{Antonio.Liotta@unibz.it}}

\maketitle              % typeset the title of the contribution

\begin{abstract}
The role of an actor in a social network is identified through a set of measures called centrality. Degree centrality, betweenness centrality, closeness centrality and clustering coefficient are the most frequently used metrics to compute the node centrality. Their computational complexity in some cases makes unfeasible, when not practically impossible, their computations. For this reason we focused on two alternative measures, WERW-Kpath and Game of Thieves, which are at the same time highly descriptive and computationally affordable. Our experiments show that a strong correlation exists between WERW-Kpath and Game of Thieves and the classical centrality measures. This may suggest the possibility of using them as useful and more economic replacements of the classical centrality measures.

% We would like to encourage you to list your keywords within
% the abstract section using the \keywords{...} command.
\end{abstract}
\section{Introduction}
\label{sec:intro}

Social network analysis (SNA) uses graph theory and networks to understand social structures and has diverse fields of application such us psychology, economics, criminology~\cite{Ficara2020, CALDERONI2020113666, Cavallaro2020, Cavallaro2021}, organizational studies and information science.

One of the most effective SNA tools to measure social interactions has been a simple graph which consists of a set of nodes (\textit{i.e.} individuals or organizations), often called actors in the SNA tradition, with edges between them, also called links or connections (\textit{i.e.} relational ties, {\em e.g.}, friendship relationships). %and with no edges connecting a node to itself.

The term node centrality indicates a family of measures aimed at characterizing the importance or the position of an actor in a network~\cite{Wasserman}. 
An actor might be important for being connected to a large number of nodes or for his power in the control over the information flowing through the network. His absence could also result in a social network made of many isolated components.
\hfill\break\indent 
Despite an abundance of methods for measuring centrality of individual nodes, there are by now only a few metrics to measure centrality of individual edges. 
De Meo {\it et al.}~\cite{MeoFFR12, MeoFFP13, de2014mixing} presented a novel measure called the $K$-path to compute link centrality. The advantage of using this metric is that it can be computed with a near-linear time algorithm called WERW-Kpath.
Most recently, Mocanu {\it et al.}~\cite{Mocanu} developed an algorithm called Game of Thieves which is able to compute actors and links centrality in a polylogarithmic time.
\hfill\break\indent 
An often asked, but rarely answered, question is: are these centrality measures correlated?~\cite{Valente2008} If there exists a high correlation between the centrality metrics, they will have a similar behavior in statistical analyses and for this reason the development of multiple measures could be redundant. If there is not high correlation, these measures are unique and they can be associated with different outcomes.
Many researchers carried out studies on the correlations between centrality measures. Valente {\it et al.}~\cite{Valente2008} investigated the correlation among the most commonly used centrality measures and they identified degree as the measure with the strongest overall correlations. Shao {\it et al.}~\cite{Shao2018} uses degree to approximate closeness, betweenness, and eigenvector in artificial and real networks. Oldham {\it et al.}~\cite{Oldham2019} used $212$ different real networks and calculated correlations between $17$ different centrality measures showing a positive correlation. %among them. 
\hfill\break\indent 
In this work, as in a previous study~\cite{ficara2020correlations}, we try to answer to a different question: are these centrality measures correlated with Game of Thieves? If so, we can use Game of Thieves thus considerably reducing the execution time in the computation of node and edge centrality in very large networks.
\hfill\break\indent 
We have done a correlation analysis using the most common types of coefficients which are Pearson, Spearman and Kendall and three kinds of artificial complex networks: scale-free, small-world and Erd\"os-Rényi random networks. %For each class, we randomly generated four or five networks which have $10,000$ nodes and from about $50,000$ to $500,000$ edges. 
\hfill\break\indent 
In a previous work~\cite{ficara2020correlations}, we analyzed the correlation coefficients among node centrality metrics when the number of nodes in both artificial and real networks increased using GoT only to measure an actor centrality. In this study, we have made an analysis on artificial networks taking into account the increase of the number of links when the network size does not change. We also used GoT to compute edges centrality comparing it with the WERW-Kpath.
\hfill\break\indent 
The paper is planned as follows. Sect.~\ref{background} presents a brief description of the main theoretical definitions of the considered centrality metrics and correlation coefficients. Sect.~\ref{res} contains a discussion of our methods and the obtained results. Sect.~\ref{conclusions} shows the conclusions and future developments of our study.

\section{Background}
\label{background}

\subsection{Basic Definitions}
\label{subsec:def}

A graph $G$ is an ordered pair $G = (N, E)$ where $N$ and $E$ are the sets of nodes and edges respectively. A graph is called undirected if edges are bidirectional.
The main node centrality measures are:
\vspace{-1mm}
\begin{itemize}
\item \textit{degree centrality} (DC)~\cite{Freeman}, which is defined as: 
\begin{equation}
DC_i = \frac{d_i}{n-1} \, ,
\end{equation}
where $d_i$ is the number of edges of a node $i$ (\textit{i.e.} the degree of $i$) and $n$ is the network size.
\item \textit{betweenness centrality} (BC)~\cite{Brandes}, which is given by the following formula: 
\begin{equation}
BC_i = \sum\limits_{h,k} \frac{v^i_{hk}}{g_{hk}} \, ,
\end{equation}
where $v^i_{hk}$ is the number of shortest paths from a node $h$ to a node $k$ by passing through $i$ and $g_{hk}$ is the total number of shortest paths from $h$ to $k$.
\item \textit{closeness centrality} (CL)~\cite{Freeman}, which is defined as:
\begin{equation}
CL_i = \frac{n}{\sum\limits_j d_{ij}} \, ,    
\end{equation}
where $d_{ij}$ is the distance between $i$ and $j$ and $n$ is the network size.
\item \textit{clustering coefficient} (CC)~\cite{Saramaki2007}, which is defined as the number of complete triangles $T_i$ in which a node $i$ participates divided by the maximum possible number of these triangles:
\begin{equation}
\label{eq:cc}
  CC_i = \frac{2T_i}{d_i (d_i - 1)} \, .
\end{equation}
\end{itemize}

A more recent centrality measure is called \textit{Game of Thieves} (GoT)~\cite{Mocanu}. 
The game proceeds in epochs. When it begins, each node has a certain number of thieves and virtual diamonds or vdiamonds. At each epoch $e$, a thief located on a node $i$ randomly picks a neighbor of $i$. He moves to this new node and, if he finds a vdiamond, he fetches it. Then, he brings the vdiamond back to his home node. 
After $T$ epochs, the centrality of each node $i$ is computed as the average number of vdiamonds present on $i$:
\begin{equation}
\Phi_T^i =  \frac{1}{T} \sum\limits_{e=0}^T \Phi_e^i \, .
\end{equation}

The centrality of each link $l$ is computed as the average number of thieves who carry a vdiamond passing through $l$:
\begin{equation}
\Psi^l_T = \frac{1}{T} \sum\limits_{e=0}^T \Psi_e^l \, .
\end{equation}

An other recent measure of link centrality for social networks is the \textit{K-path} which is defined as:
\begin{equation}
L^k(l) = \sum\limits_i \frac{\sigma_i^k(l)}{\sigma_i^k} \, ,
\end{equation}
where $\sigma_i^k(l)$ is the number of $k$-paths originating from a node $i$ and traversing the link $l$ and $\sigma_i^k$ is the number of $k$-paths which originate from $i$. 
A near linear time algorithm called WERW-Kpath (Weighted Edge Random Walks – $K$ Path)~\cite{MeoFFR12, MeoFFP13, de2014mixing} is able to compute this centrality index. 

\begin{table}[htp]
\caption{\textbf{Centrality measures.} $n$ is the cardinality of $N$ and $m$ is the cardinality of $E$ in a graph $G(N,E)$.}
\begin{center}
\begin{tabular}{r@{\quad}r@{\qquad}rr}
\hline\rule{0pt}{12pt}
Measure & Centrality & Computational Complexity \\
\hline\rule{0pt}{12pt}
DC & Nodes & $O(m)$ &\\[4pt]
BC & Nodes & $O(mn)$ or $O(n^3)$ &\\[4pt]
CL & Nodes & $O(n^3)$ &\\[4pt]
CC & Nodes & $O(n^2)$ &\\[4pt]
GoT & Nodes and Edges & $O(log^2n)$ or $O(log^3n)$ &\\[4pt]
WERW-Kpath & Edges & $O(km)$ &\\[4pt]
\hline
\end{tabular}
\end{center}
\label{tab:table1}
\end{table}
%\vspace{-4mm}
Table~\ref{tab:table1} shows how GoT represents a great step forward in terms of computational complexity with respect to classical algorithms of node centrality such us BC, CL and CC which have a quadratic computational complexity. The DC has a linear time complexity but GoT still remains a better option because it is also able to compute the centrality of edges.

\subsection{Correlation Coefficients}
\label{subsec:corr}

Given two random variables $a$ and $b$, the \textit{Pearson's $r$ correlation coefficient}~\cite{Chen} is defined as:
\begin{equation}
r = \frac{cov(a,b)}{\sigma_a \sigma_b} \, ,
\end{equation}
where $cov(a,b)$ is the covariance of $a$ and $b$ and $\sigma_a \sigma_b$ is the product of their standard deviations.

The \textit{Spearman's $\rho$ rank correlation coefficient}~\cite{Spearman} is defined as the Pearson's $r$ between the rank values of the two variables. For a sample of size $s$, the $s$ raw scores $a$ and $b$ are converted to ranks $rg_a$ and $rg_b$, and $\rho$ is computed as:
\begin{equation}
\rho = \frac{cov(rg_a,rg_b)}{\sigma_{rg_a} \sigma_{rg_b}} \, ,
\end{equation}
where $cov(rg_a,rg_b)$ is the covariance of the rank variables and $\sigma_{rg_a} \sigma_{rg_b}$ are the standard deviations of the rank variables.

Given two samples $a$ and $b$, where each sample size is $s$, {\it Kendall's $\tau$ rank correlation coefficient}~\cite{Kendall} is defined according to the following formula: 
\begin{equation}
\tau = \frac{s_c - s_d}{\frac{1}{2}s(s - 1)} \, ,
\end{equation}
where $s_c$ and $s_d$ are the numbers of concordant and discordant pairs respectively and $\frac{1}{2}s(s-1)$ is the total number of pairings with $a$ and $b$.

\section{Methodology and Results}
\label{res}

We investigated the correlations among the node and edge centrality measures, described in Subsect.~\ref{subsec:def}, in different kinds of artificial networks. The network models include the scale-free (SF)~\cite{Barabasi}, the small-world (SW)~\cite{NEWMAN1999341} and the Erd\"os-Rényi (ER)~\cite{erdos59a} random networks. For each class, we randomly generated undirected and unweighted networks with $10,000$ nodes and between $5,000$ and $50,000$ edges. 

We used  the model proposed by Holme and Kim~\cite{Holme2002} to generate the SF networks. In each experiment, we chose the number of nodes $n = 10,000$, we added $m = \{5, 15, 25, 35, 50\}$ random edges for each new node $i$, and we picked a probability $p = 0.3$ of adding a triangle after adding a random edge.

SW graphs are generated using Newman-Watts-Strogatz (NWS) small-world model~\cite{NEWMAN1999341}. In each experiment, we chose the number of nodes $n = 10,000$,  $k = \{6, 18, 32, 64\}$ neighbors with which connect each node $i$ in the ring topology, and a probability $p=0.6$ of rewiring each edge.

\begin{figure}
\includegraphics[width=1.02\textwidth]{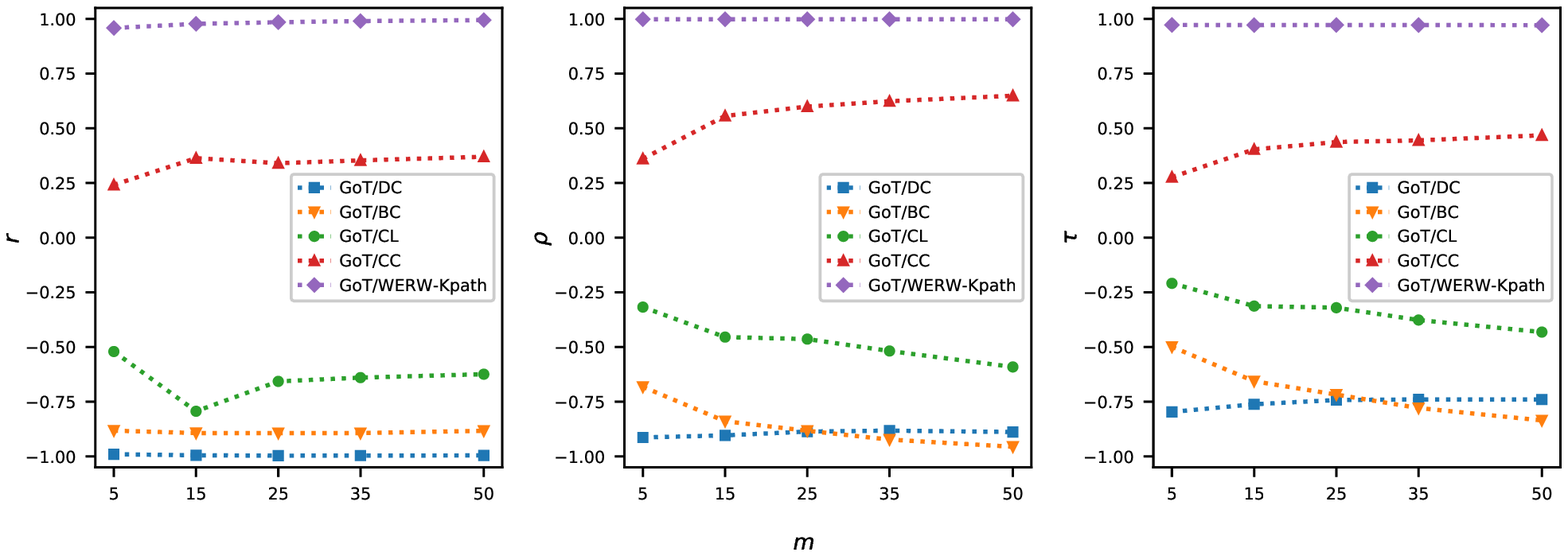}
\includegraphics[width=1.02\textwidth]{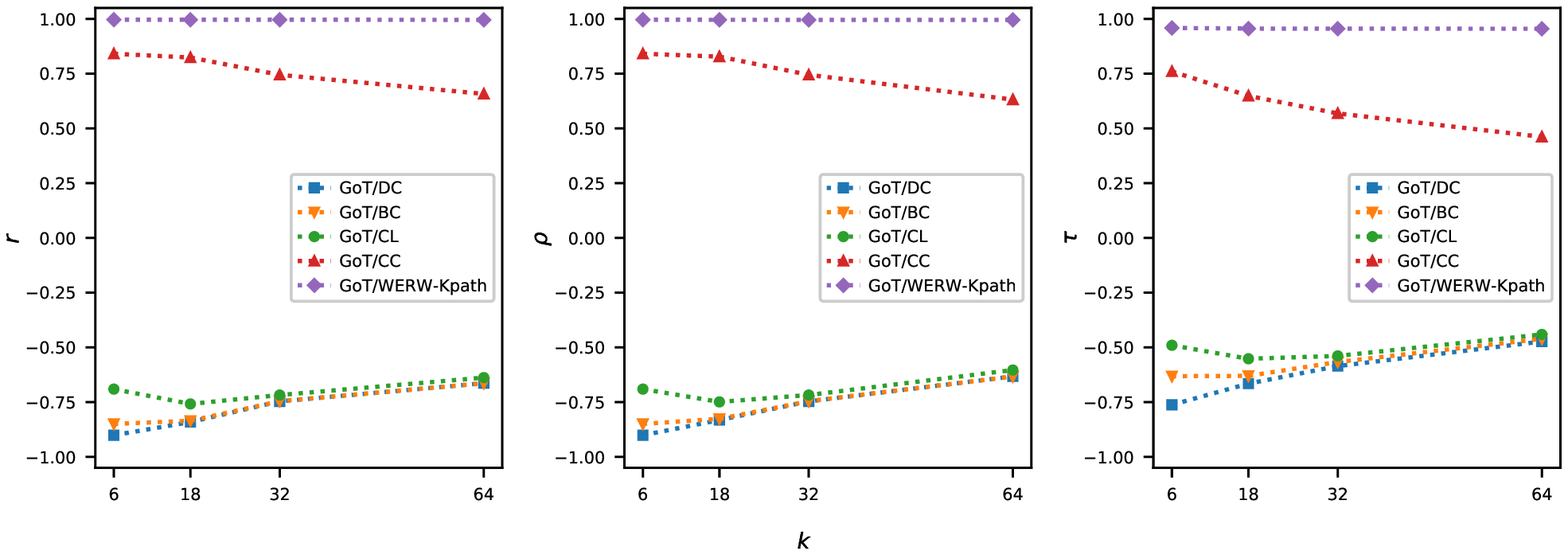}
\includegraphics[width=1.02\textwidth]{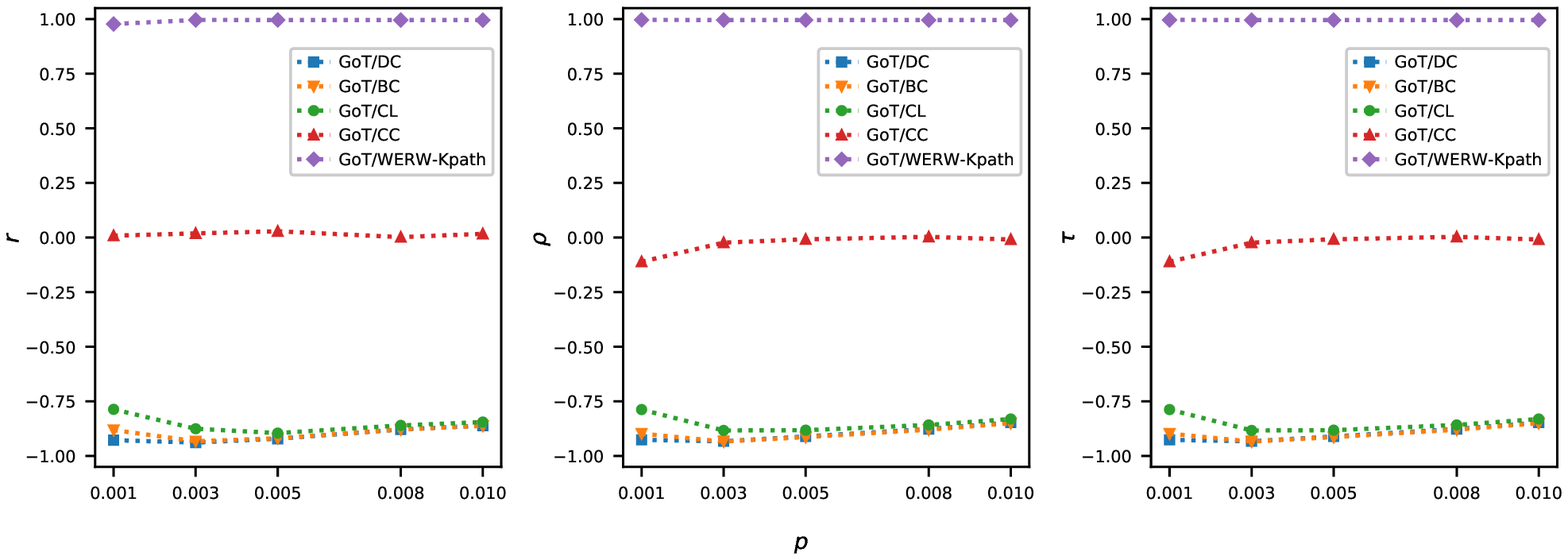}

\caption{Pearson's $r$ \textbf{(a)}, Spearman's $\rho$ \textbf{(b)} and Kendall's $\tau$ \textbf{(c)} correlation coefficients.}
\label{fig:SF}
\end{figure}

ER networks are generated choosing the network size $n = 10,000$, and for edge creation the probability values $p = \{0.001, 0.003, 0.005, 0.010\}$.
\hfill\break\indent 
For the implementation of the artificial networks and the centrality metrics such as DC, BC, CL, and CC, we used Python and NetworkX module~\cite{SciPyProceedings11}. To run the WERW-Kpath\footnote[1]{Available at  \url{http://www.emilio.ferrara.name/code/werw-kpath/}}~\cite{MeoFFR12, MeoFFP13, de2014mixing} we set the value of the random walk to $k=10$.  For GoT\footnote[2]{Available at \url{http://github.com/dcmocanu/centrality-metrics-complex-networks}} we set $1$ thief per node and the initial amount of vdiamonds per node equal to the network size. We let GoT to run for $T=\log^3 n$ epochs. 
\hfill\break\indent 
In Fig.~\ref{fig:SF} are shown the results of the Pearson's $r$ correlation coefficient (left column), the Spearman's $\rho$ (middle column) and the Kendall's $\tau$ rank correlation coefficients (right column) for SF networks (upper row), for SW networks (middle row) and ER random networks (lower row).
\hfill\break\indent 
In SF networks (see Fig.~\ref{fig:SF} upper row), GoT and DC have the strongest negative correlation. GoT exhibits a strong negative correlation with the BC and slightly weaker with the CL. There is a positive correlation between GoT and the CC which becomes stronger for the rank correlation coefficients. In particular, we can observe that the Sperarman's $\rho$ is always larger than the Kendall's $\tau$.
Except for the DC, we can observe a small deviation of the rank correlation coefficients when the number of edges in the analyzed networks is smaller which is not visible anymore when edges grow big enough.
\hfill\break\indent 
In SW networks (see Fig.~\ref{fig:SF} middle row), there is an almost identical strong negative correlation among GoT and DC, BC and CL. We can also still observe a strong positive correlation between GoT and the CC. All of these correlations become weaker when the number of edges grows.
\hfill\break\indent 
In ER networks (see Fig.~\ref{fig:SF} lower row), the correlation among GoT and DC, BC and CL is the same as in SW networks. But, in this kind of networks, there is no correlation between GoT and the CC even if we can observe a very little deviation when the number of edges is small.
\hfill\break\indent 
In all networks, we can observe a strong positive correlation among GoT and WERW-Kpath which remains constant when the number of edges grows.

\section{Conclusions}
\label{conclusions}

In this work we used artificial networks, {\em i.e.} SF, SW and ER networks, to examine the correlation between well known and more recently proposed centrality measures. 
A strong correlation implies the possibility of approximating the metric with the highest computational complexity using the other. 
We have done an analysis observing the Pearson, Spearman and Kendall correlation coefficients on artificial networks by changing the amount of links given a fixed network size. 
Our results show a strong correlation among GoT and the most known centrality algorithms. For this reason, it can substitute them in the computation of node centrality in large networks. GoT can also replace the CC as a property of SW and SF networks.
\hfill\break\indent 
A future development of our work will be to focus more on the application of WERW-Kpath, changing the values of $k$, and GoT to the case of edge centrality. 
The two measures seems to be equivalent and it could be also interest to make a deeper analysis on their computational costs on large networks.



\begin{thebibliography}{10}
\providecommand{\url}[1]{{#1}}
\providecommand{\urlprefix}{}
\expandafter\ifx\csname urlstyle\endcsname\relax
  \providecommand{\doi}[1]{doi:~\discretionary{}{}{}#1}\else
  \providecommand{\doi}{doi:~\discretionary{}{}{}\begingroup
  \urlstyle{rm}\Url}\fi

\bibitem{Ficara2020}
Ficara, A., Cavallaro, L., De~Meo, P., Fiumara, G., Catanese, S., Bagdasar, O., Liotta, A.: Social network analysis of {Sicilian Mafia} interconnections.
\newblock In: H.~Cherifi, S.~Gaito, J.F. Mendes, E.~Moro, L.M. Rocha (eds.) Complex Networks and Their Applications VIII, pp. 440--450. Springer International Publishing, Cham (2020).
\newblock \doi{10.1007/978-3-030-36683-4_36}

\bibitem{CALDERONI2020113666}
Calderoni, F., Catanese, S., {De Meo}, P., Ficara, A., Fiumara, G.: Robust link prediction in criminal networks: A case study of the {Sicilian Mafia}.
\newblock Expert Systems with Applications \textbf{161}, 113,666 (2020).
\newblock \doi{10.1016/j.eswa.2020.113666}

\bibitem{Cavallaro2020}
Cavallaro, L., Ficara, A., {De Meo}, P., Fiumara, G., Catanese, S., Bagdasar, O., Song, W., Liotta, A.: Disrupting resilient criminal networks through data analysis: The case of {Sicilian Mafia}.
\newblock PLOS ONE \textbf{15}(8), 1--22 (2020).
\newblock \doi{10.1371/journal.pone.0236476}

\bibitem{Cavallaro2021}
Cavallaro, L., Ficara, A., Curreri, F., Fiumara, G., De~Meo, P., Bagdasar, O., Liotta, A.: Graph comparison and artificial models for simulating real criminal networks.
\newblock In: R.M. Benito, C.~Cherifi, H.~Cherifi, E.~Moro, L.M. Rocha, M.~Sales-Pardo (eds.) Complex Networks and Their Applications IX, pp. 286--297. Springer International Publishing, Cham (2021)
\newblock \doi{10.1007/978-3-030-65351-4_23}

\bibitem{Wasserman}
Wasserman, S., Faust, K., Granovetter, M., Iacobucci, D.: Social Network Analysis: Methods and Applications.
\newblock Structural Analysis in the Social Sciences. Cambridge University Press (1994)

\bibitem{MeoFFR12}
{De Meo}, P., Ferrara, E., Fiumara, G., Ricciardello, A.: A novel measure of edge centrality in social networks.
\newblock Knowledge Based Systems \textbf{30}, 136--150 (2012).
\newblock \doi{10.1016/j.knosys.2012.01.007}

\bibitem{MeoFFP13}
{De Meo}, P., Ferrara, E., Fiumara, G., Provetti, A.: Enhancing community detection using a network weighting strategy.
\newblock Information Sciences \textbf{222}, 648--668 (2013).
\newblock \doi{10.1016/j.ins.2012.08.001}

\bibitem{de2014mixing}
{De Meo}, P., Ferrara, E., Fiumara, G., Provetti, A.: Mixing local and global information for community detection in large networks.
\newblock Journal of Computer and System Sciences \textbf{80}(1), 72--87 (2014).
\newblock \doi{10.1016/j.jcss.2013.03.012}

\bibitem{Mocanu}
Mocanu, D.C., Exarchakos, G., Liotta, A.: Decentralized dynamic understanding of hidden relations in complex networks.
\newblock Scientific Reports \textbf{8}(1), 1571 (2018).
\newblock \doi{10.1038/s41598-018-19356-4}

\bibitem{Valente2008}
Valente, T.W., Coronges, K., Lakon, C., Costenbader, E.: How correlated are network centrality measures?
\newblock Connections (Toronto, Ont.) \textbf{28}(1), 16--26 (2008)

\bibitem{Shao2018}
Shao, C., Cui, P., Xun, P., Peng, Y., Jiang, X.: Rank correlation between centrality metrics in complex networks: An empirical study.
\newblock Open Physics \textbf{16}(1), 1009--1023 (2018).
\newblock \doi{10.1515/phys-2018-0122}

\bibitem{Oldham2019}
Oldham, S., Fulcher, B., Parkes, L., Arnatkeviciute, A., Suo, C., Fornito, A.: Consistency and differences between centrality measures across distinct classes of networks.
\newblock PLOS ONE \textbf{14}(7), 1--23 (2019).
\newblock \doi{10.1371/journal.pone.0220061}

\bibitem{ficara2020correlations}
Ficara, A., Fiumara, G., De~Meo, P., Liotta, A.: Correlations among {Game of Thieves} and other centrality measures in complex networks.
\newblock In: G.~Fortino, A.~Liotta, R.~Gravina, A.~Longheu (eds.) Data Science and Internet of Things. Springer International Publishing (2021).
\newblock \doi{10.1007/978-3-030-67197-6_3}

\bibitem{Freeman}
Freeman, L.C.: Centrality in social networks conceptual clarification.
\newblock Social Networks \textbf{1}(3), 215--239 (1978).
\newblock \doi{10.1016/0378-8733(78)90021-7}

\bibitem{Brandes}
Brandes, U.: On variants of shortest-path betweenness centrality and their generic computation.
\newblock Social Networks \textbf{30}(2), 136--145 (2008).
\newblock \doi{10.1016/j.socnet.2007.11.001}

\bibitem{Saramaki2007}
Saram\"aki, J., Kivel\"a, M., Onnela, J.P., Kaski, K., Kert\'esz, J.: Generalizations of the clustering coefficient to weighted complex networks.
\newblock Phys. Rev. E \textbf{75}, 027,105 (2007).
\newblock \doi{10.1103/PhysRevE.75.027105}

\bibitem{Chen}
Chen, P., Popovich, P.: Correlation: parametric and nonparametric measures.
\newblock Sage university papers series. No. 07-139. Sage Publications (2002)

\bibitem{Spearman}
Spearman, C.: General intelligence, objectively determined and measured.
\newblock The American Journal of Psychology \textbf{15}(2), 201--292 (1904).
\newblock \doi{10.2307/1412107}

\bibitem{Kendall}
Kendall, M., Gibbons, J.: Rank Correlation Methods.
\newblock Charles Griffin Book. E. Arnold (1990)

\bibitem{Barabasi}
Barab{\'a}si, A.L., Albert, R.: Emergence of scaling in random networks.
\newblock Science \textbf{286}(5439), 509--512 (1999).
\newblock \doi{10.1126/science.286.5439.509}

\bibitem{NEWMAN1999341}
Newman, M., Watts, D.: Renormalization group analysis of the small-world network model.
\newblock Physics Letters A \textbf{263}(4), 341--346 (1999).
\newblock \doi{10.1016/S0375-9601(99)00757-4}

\bibitem{erdos59a}
Erd\"{o}s, P., R\'{e}nyi, A.: On Random Graphs I.
\newblock Publicationes Mathematicae Debrecen \textbf{6}, 290 (1959)

\bibitem{Holme2002}
Holme, P., Kim, B.J.: Growing scale-free networks with tunable clustering.
\newblock Phys. Rev. E \textbf{65}, 026,107 (2002).
\newblock \doi{10.1103/PhysRevE.65.026107}

\bibitem{SciPyProceedings11}
Hagberg, A.A., Schult, D.A., Swart, P.J.: Exploring network structure, dynamics, and function using networkx.
\newblock In: G.~Varoquaux, T.~Vaught, J.~Millman (eds.) Proceedings of the 7th Python in Science Conference, pp. 11--15. Pasadena, CA USA (2008)

\end{thebibliography}
\end{document}